\documentclass{svmult}
\usepackage{amsmath,amssymb}
\usepackage{graphicx}

\mainmatter 
\begin{document}

\title{Creation of Clusters via a Thermal Quench}
\author{Yossi Farjoun} \institute{Department of Mathematics,
  MIT}
 
\maketitle
\begin{abstract}
  The nucleation and growth of clusters in a progressively cooled
  vapor is studied.  
  The chemical-potential of the vapor increases, resulting in a
  rapidly increasing nucleation rate.  
  The growth of the newly created clusters depletes monomers, and counters the
  increase in chemical-potential.  
  Eventually, the chemical potential reaches a maximum and begins
  to decrease. 
  Shortly thereafter the nucleation of new clusters
  effectively ceases.  
  Assuming a slow quench rate, asymptotic methods are used to convert
  the non-linear advection equation of the cluster-size distribution
  into a fourth-order differential equation, which is solved numerically. 
  The distribution of cluster-sizes that emerges from this
  \emph{creation} era of the quench process, and the total amount of
  clusters generated are found.   
\end{abstract}
\section{Introduction}
While studying a simplified model of nucleation the temperature is often
\emph{assumed} to be held constant (see, e.g., the review papers by
Wu \cite{Wu96} and Oxtoby \cite{Oxtoby92}). 
However, time dependence of the temperature can play an important role
in the aggregation process.  
In this paper we study the effect that a thermal quench has on aggregation. 
Together with J.~Neu, we previously studied
\cite{FN:Creation:08} a similar problem with \emph{constant} temperature.  
For a less terse literature summary, the reader is referred to the
review articles cited above.

Our system comprises a dilute, condensable vapor in a carrier gas.
Starting from a temperature corresponding to zero chemical-potential,
the system is cooled uniformly at a constant rate.  We assume a quench
rate that is slow relative to the molecular timescale. This implies
that the maximal chemical-potential is small and therefore the nucleation
rate and cluster growth can be assumed to follow the Zeldovich
formula \cite{ZELD43} and the Becker-D\"oring (BD) equations \cite{BD35}
respectively.  We use the BD equations and not diffusion limited
growth, as per Lifshitz-Slyozov \cite{LS61}, since the final size of the
clusters is relatively small.  To find the value of the
chemical-potential we use conservation of particles and the
Clasius-Clapyron relation together with the ideal-gas law.  To
simplify the model and calculation, we assume that the latent heat of
evaporation is large relative to the thermal energy, $k_BT$.

Initially, there are very few clusters and so the chemical-potential
increases as the temperature drops (through the resulting reduction of
the equilibrium monomer density).  As the chemical-potential increases,
so does the nucleation rate.  Eventually, enough clusters have been
created that their growth causes a large enough combined drain on the
monomer density so that the increase in chemical-potential is stopped.  
After this, the nucleation rate drops quickly and approaches zero.

The paper is organized as follows: In Section \ref{sec:model} we
present a short derivation of the aggregation model we use. In Section
\ref{sec:solution} we solve the resulting non-linear advection PDE using an asymptotic
approximation, the method of characteristics, and a numerical solver. 
\subsection{Assumptions}
Throughout the paper we make the following assumptions.
\begin{description}
\item[1.] Nucleation occurs at the Zeldovich rate.
\item[2.] The clusters that form are small, and grow according
  to BD dynamics.
\item[3.] The process is spatially uniform.
\item[4.] The only interaction between clusters is via the chemical-potential.
\item[5.] The temperature is not affected from the condensation of
  clusters.
\item[6.] The carrier gas and the condensable gas are ideal gasses.
\end{description}
\section{The Model}
\label{sec:model}
We briefly derive the constituent equations of the quenching process.
We assume basic familiarity with standard nucleation, BD growth, and
the monomer conservation argument that leads to the determination of
the chemical-potential from the distribution of cluster sizes.  For
these we follow the notation in Wu's review article \cite{Wu96}.
First, we derive the relationship between the chemical-potential and
the undercooling.

\subsection{Chemical-potential and the Undercooling}
The chemical-potential $\eta$ is the free energy of a monomer in
condensed liquid relative to that in vapor:
\begin{equation}
  \eta = k_BT\log\frac{c}{c_e}.\label{eq:eta:def}
\end{equation}
Here, $k_B$ is the Bolzmann constant, $T$ is the temperature, $c$ is
the ambient monomer concentration, and $c_e$ is the equilibrium
monomer concentration.

The dependence of $c_e$ on the temperature can be found through the
Clausius-Clyperon relation and the ideal-gas law as in \cite{CR89}. 
\begin{equation}
  \frac{c_e}{c_0} = \frac{T_0}{T}
  \E^{\frac{\Lambda}{k_B T_0}-\frac{\Lambda}{k_BT}}, \label{eq:c_e:def}
\end{equation}
where $\Lambda$ is the latent-energy of condensation (per monomer.)
In \eqref{eq:c_e:def}, $T_0$ and $c_0$ are the initial equilibrium
temperature and monomer-concentration.  That is, at concentration
$c_0$ and temperature $T_0$ the vapor phase is in equilibrium with the
liquid phase.  Thus, setting $T=T_0$ gives $c_e=c_0$.

We assume that $\Lambda\gg k_BT_0$ and define $\varepsilon\equiv\frac\Lambda{k_BT_0}$.
The non-dimensional versions of $c$, $T$, and $\eta$ are:
\begin{equation}
  \tilde c = \frac{c}{c_0},\qquad \tilde T = \frac{T}{T_0},\qquad
  \tilde \eta = \frac{\eta}{k_BT_0}.
\end{equation}
Equations \eqref{eq:eta:def} and \eqref{eq:c_e:def} now are as follows
\begin{equation}
  \tilde\eta = \tilde T \log\frac{\tilde c}{\tilde c_e},\qquad \tilde
  c_e={\textstyle\frac{1}{\tilde T}}
  \E^{\frac{1}{\varepsilon}-\frac{1}{\varepsilon \tilde T}}.
\end{equation}
The undercooling is defined as $\tau \equiv \frac{1-\tilde
  T}{\varepsilon}$.

Using the above definition of $tau$ we find an asymptotic approximation of
$\tilde c_e$ for small $\varepsilon$:
\begin{equation}
  \tilde c_e =  \E^{-\tau} +O(\varepsilon\tau),\label{eq:c_e:asymp}
\end{equation}
and the leading order approximation of $\tilde\eta$:
\begin{align}
  \tilde\eta  &\approx \log \tilde c + \tau, \quad \text{ for }
  \varepsilon\ll1.\label{eq:asymptotic:eta}
\end{align}

\subsection{The Zeldovich nucleation rate and the growth rate of clusters}
Two important components of our model are the rate at which new
clusters come into existence, and the rate at which existing clusters
grow.  The BD equations of growth specify that the size $n$ of a
cluster follows the growth ``law'' for clusters much larger than the
critical size:
\begin{equation}
  \dot n = \omega \tilde\eta n^\frac23.\label{eq:n_dot}
\end{equation} 
Where $\eta$ is as in \eqref{eq:eta:def}, and $\omega$ is a ``escape
rate'' constant so that $\omega n^{2/3}$ is the rate at which monomers
leave the cluster.  

The Zeldovich formula give the nucleation rate of new clusters:
\begin{equation}
  j=\omega c_0 \tilde
  c_e\sqrt{\frac{\sigma}{6\pi}}\E^{-\frac{T_0}{T}\frac{\sigma^3}{2\tilde\eta^2}}.
\end{equation}
Where $\sigma$ is the ``surface energy'' constant of a cluster, 
\begin{equation}
  \frac{\text{surface energy}}{k_BT_0}=\frac32\sigma n^{\frac23}.\label{eq:sig:def}
\end{equation}
Using the definition of $\tau$ and \eqref{eq:c_e:asymp} we have
\begin{equation}
  j=\omega
  c_0\sqrt{\frac{\sigma}{6\pi}}\E^{-\frac{1}{1-\varepsilon
      \tau}\frac{\sigma^3}{2\tilde\eta^2}-\tau}
  \label{eq:nucleation_rate}
\end{equation}

\subsection{Growth dynamics via advection PDE}
We are interested in finding the evolution of the density of
cluster sizes.  Let $r(n,\, t)$ be the density of clusters of size $n$
at time $t$ (also referred to as the ``distribution of cluster-sizes'').
From equation \eqref{eq:n_dot} we derive an advection PDE for the
distribution of clusters in the space of their size,
\begin{equation}
  \partial_t r + \omega \tilde\eta\partial_n\left(n^{\frac32}r\right) =0,\qquad\text{
    in } n>0.
\end{equation}
The flux of clusters $\omega\tilde\eta n^{\frac23} r$ asymptotes
to the Zeldovich rate as $n\rightarrow0$, thus
\begin{equation}
  \omega\tilde\eta n^\frac23 r \rightarrow j = \omega c_0
  \sqrt{\frac{\sigma}{6\pi}}\E^{-\frac{\sigma^3}{2\tilde\eta^2}-\tau}
  \text{ as }n\rightarrow0.\label{eq:BC:limit}
\end{equation}

\subsection{Determination of the chemical-potential}
The chemical-potential can be inferred from the distribution of
cluster sizes and the initial concentration $c_0$.  
\begin{align}
  c + \int_0^\infty n r(n,\,t)\, dn &=c_0.  \intertext{For
    small values of $\eta$ we have from \eqref{eq:asymptotic:eta} that
    $c=c_0 e^{\tilde\eta-\tau}$. Thus} \E^{\tilde\eta-\tau} +
  \int_0^\infty n \tilde r(n,\,t)\, dn &=1, \text{ where  }\tilde r = \frac{r}{c_0}.
\end{align}

\section{The Mathematical Problem}
\label{sec:solution}
We now drop all tildes and refer only to non-dimensional variables.
The mathematical problem is therefore,
\begin{align}
  \partial_t r + \eta \partial_n(n^\frac23  r)&=0,\label{eq:PDE:scaled:1}\\
  \eta n^\frac23r\rightarrow
  \sqrt{\frac{\sigma}{6\pi}}\E^{-\frac{\sigma^3}{2\eta^2}-\tau},\quad
  \text{ as } n&\rightarrow0,\label{eq:BC:scaled:1}\\
  \E^{\eta-\tau}+\int_0^\infty nr\,dr
  &=1.\label{eq:conservation:scaled:1}
\end{align}
Here, time $t$ is non-dimensionalized with the scaling unit $1/\omega$.

In the current paper we consider an undercooling which
increases linearly with time $t$:
\begin{equation}
  \tau = \Omega (t+t_0).
\end{equation}
The parameter $\Omega$ is externally specified, and
$t_0$ is a time-lag needed so that the ``interesting'' behavior
happens near $t=0$. 

\subsection{Asymptotic Solution of the Creation Era}

The equations needed to find the relevant scales of the
creation era are mostly straightforward dominant balances of 
equation (\ref{eq:PDE:scaled:1}--\ref{eq:conservation:scaled:1}). 
There is one that is not: The change in chemical-potential must be
such that the reduction in the nucleation rate is comparable to the
nucleation rate itself. 
This implies that the change in chemical-potential is small (relative
to $\eta$ itself) and we use $\delta\eta\equiv\eta(0)-\eta(t)$ to
follow the \emph{change} in the chemical potential.
To save space, we omit the derivation and proceed to the resulting scales. 
They are
\begin{equation*}
[\delta\eta] = \left(\frac{\Omega
      t_0}{\sigma}\right)^3 \quad [n]=\Omega^3 t_0^3\left(\frac{1}{\sigma^3 E}\right)^\frac34 \quad
[t]={\left(\frac{1}{\sigma^3 E}\right)}^\frac14\quad
[r]={\left(\frac{\sigma E}{\Omega^2 t_0^2}\right)}^\frac32.
\end{equation*}
While $t_0$ and $\Omega$ are connected by $\Omega^2 t_0^3
E^\frac14=\sigma^\frac94$.
We introduced $E$, a measure of nucleation rate: $E\equiv \sqrt{\frac{\sigma}{6\pi}}\E^{-\frac{\sigma^3}{2(t_0\Omega)^2}-\Omega t_0}$.
Using these scales results in the following system for $r(n,\,t)$ and $\eta$:
\begin{align}
  \partial_t r + \partial_n(n^\frac23 r)=0,\label{eq:PDE:scaled}\\
  n^\frac23r\rightarrow \E^{\delta\eta} \text{ as } n\rightarrow0,\label{eq:PDE_BC:scaled}\\
  \delta\eta(t) - t +\int_0^\infty nr\,dr =0.
\label{eq:PDE_CONSERV}
\end{align}
Since  $n^\frac23 r$ is constant along the level curves of $3 n^{\frac13}-t$,
 we can write
\begin{equation}
  r(n,t) = n^{-\frac23}{\E^{\delta\eta(t-3n^{\frac13})}}.
\end{equation}
Substituting this into \eqref{eq:PDE_CONSERV}
yields an integral equation for $\delta\eta$, and 
using the change of variable $t' = t-3n^{\frac13}$ we arrive at
\begin{equation}
  \delta\eta(t) - t +\int_{-\infty}^t {\left(\frac{3}{t-t'}\right)}^3 \E^{\delta\eta(t')}\,dt' =0.
\end{equation}
From here we derive a fourth-order ordinary differential equation (ODE) for $\delta\eta$, by
differentiating 4 times:
\begin{equation}
  \ddddot{\delta\eta}=-\frac29 \E^{\delta\eta}, \qquad\text{ where }
  \delta\eta(t)\sim t, \text{ as } t\rightarrow -\infty.
\end{equation}
To solve this ODE (numerically), we start with \mbox{$\delta\eta(t_i)= t_i$},
\mbox{$\dot{\delta\eta}(t_i)=1$}, \mbox{$\ddot{\delta\eta}(t_i)=0$},
\mbox{$\dddot{\delta\eta}(t_i)=0$}, and $t_i=-10$.

\begin{figure}[t]
\resizebox{\textwidth}{!}{\includegraphics{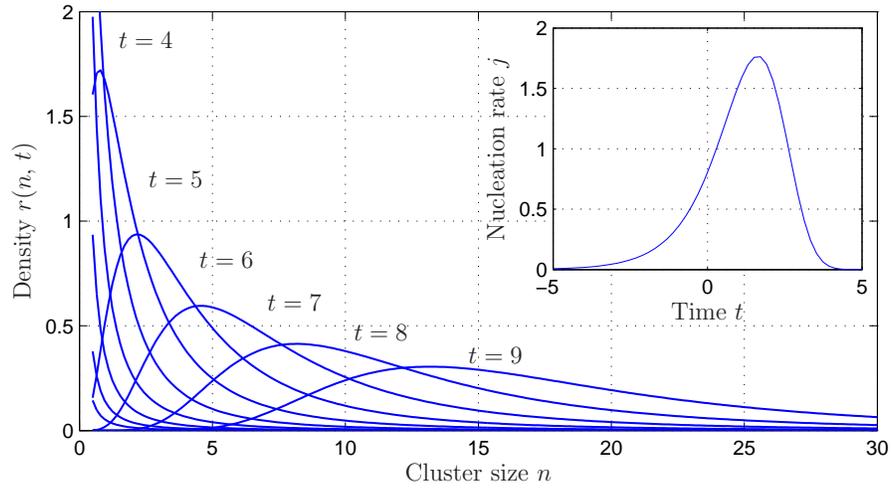}}
  \caption{The cluster size distribution $r(n,t)$ for $t$ ranging from
    $0$ to $9$ and, inset, the nucleation rate as a function of (shifted and scaled) time.} 
\label{fig:density}
\end{figure}

By integrating $\E^{\delta\eta}$  we get the
(scaled) density of clusters $\rho$ that were formed, $\rho \approx 5.1$\@.
The resulting density is shown in Fig.~\ref{fig:density}.
\section{Conclusions}
Thermal quench is a standard trigger for nucleation.  However,
normally the quench process itself is ignored and only the outcome
(i.e., a super-saturated monomer solution) is considered.  In this
paper we have shown that during a slow quench, enough clusters
nucleate so that no more nucleate after the quench.  Presumably, the
clusters that have nucleated simply grow after the quench and
eventually coarsen.  
\bibliographystyle{ECMIplain}
\bibliography{general}
\end{document}